\documentclass[a4paper]{spie}
\usepackage{graphicx}
\title{A very fast iterative algorithm for TV-regularized image reconstruction with applications to low-dose and few-view CT}
\author{Hiroyuki Kudo\supit{a,b}, Fukashi Yamazaki\supit{a}, Takuya Nemoto\supit{a}, and Keita Takaki\supit{a}
\skiplinehalf
\supit{a}Faculty of Engineering, Information and Systems, University of Tsukuba, Tennoudai 1-1-1, Tsukuba 305-8573, Japan \\
\supit{b}JST-ERATO Momose Quantum-Beam Phase Imaging Project, Katahira, Aoba-ku, Sendai 980-8577, Japan
}
\authorinfo{Further author information: (Send correspondence to H.K.)\\H.K.: E-mail: kudo@cs.tsukuba.ac.jp}
\begin{document}
\maketitle
\begin{abstract}
This paper concerns iterative reconstruction for low-dose and few-view CT by minimizing a data-fidelity term regularized with the Total Variation (TV) penalty. We propose a very fast iterative algorithm to solve this problem. The algorithm derivation is outlined as follows. First, the original minimization problem is reformulated into the saddle point (primal-dual) problem by using the Lagrangian duality, to which we apply the first-order primal-dual iterative methods. Second, we precondition the iteration formula using the ramp filter of Filtered Backprojection (FBP) reconstruction algorithm in such a way that the problem solution is not altered. The resulting algorithm resembles the structure of so-called iterative FBP algorithm, and it converges to the exact minimizer of cost function very fast.
\end{abstract}
\section{INTRODUCTION}
\par Within the famous Total Variation (TV) regularization framework, image reconstruction in low-dose CT is formulated as minimizing the cost function expressed as $f(\vec{x})=\beta\parallel\vec{x}\parallel_{\rm TV}+\parallel A\vec{x}-\vec{b}\parallel_W^2$ subject to $\vec{x}\geq 0$, and that in few-view CT is formulated as $\min_{\vec{x}} \parallel\vec{x}\parallel_{\rm TV}\ {\rm subject}\ {\rm to}\ A\vec{x}=\vec{b},\ \vec{x}\geq 0$, where $\parallel\vec{x}\parallel_{\rm TV}$ is the TV norm which aims at regularizing the ill-conditioned reconstruction problem. A variety of iterative algorithms have been proposed for solving the above problems in CT reconstruction fields based on modifying classical iterative algorithms such as the gradient method, ART (Algebraic Reconstruction Technique), SIRT (Simultaneous Iterative Reconstruction Technique), and SART (Simultaneous Algebraic Reconstruction Technique), often combined with the Ordered-Subsets (OS) technique [1]-[10] (and many others). However, it is fair to say that there exist very few iterative algorithms in engineering literatures, which not only converge fast but also can exactly solve the above TV-regularized problems. It is known that, from mathematical point of view, this difficulty partly comes from the non-differentiability of TV penalty. However, in applied mathematics fields, the research progress on this topic is very fast so that many nice algorithms to handle the TV regularization have been developed since 2010. Therefore, we believe that it would be time for CT engineers like us to develop and use more rigorous iterative algorithms for the TV-regularized image reconstructions.
\par In this paper, we propose a very fast iterative algorithm which can be applied to the above two typical TV-regularized image reconstructions in a unified way. The algorithm derivation is outlined as follows. First, the original minimization problem is reformulated into the standard saddle point (primal-dual) problem by using the Lagrangian duality, to which we apply the Alternating Projection Proximal (APP) algorithm which belongs to a class of first-order primal-dual methods including Chambolle-Pock algorithm, Generalized Iterative Soft-Thresholding (GIST) algorithm, and Alternating Extragradient (AE) algorithm [11]-[14]. However, the resulting algorithm converges very slowly, mainly because its overall structure is same as the simultaneous iterative reconstruction methods such as the classical SIRT and SART algorithms. To overcome this drawback, we precondition the iteration formula using the ramp filter of Filtered Backprojection (FBP) reconstruction algorithm in such a way that the solution to the preconditioned iteration perfectly coincides with the solution to the original problem. The final algorithm can be interpreted as the first-order primal-dual method accelerated by the FBP-type preconditioning using the ramp filter. Unlike the famous known FBP-based acceleration technique called the iterative FBP algorithm [15],[16], for both the above two different formulations (low-dose CT and few-view CT), the proposed algorithm converges to the solution exactly minimizing the cost function. We evaluate this algorithm for both the low-dose CT case and the few-view CT case with 32 projection data. In the low-dose CT experiment, the proposed algorithm converged with approximately 10 iterations to the almost same image as the GIST algorithm with 1,000 iterations. In the few-view CT experiment, the proposed algorithm required only 3 iterations to reach to the almost same image as the Chambolle-Pock algorithm with 1,000 iterations. These accelerations were achieved by the introduction of FBP-type preconditioning in the primal-dual space (in the saddle point formulation), which has not been investigated yet and is an original contribution of this paper.

\section{PROPOSED ALGORITHM FOR LOW-DOSE CT RECONSTRUCTION}
\subsection{Problem Formulation and Brief Review of Existing Algorithms}
\par In this section, we formulate image reconstruction in the low-dose CT and review typical existing iterative algorithms. We denote an image by a $J$-dimensional vector $\vec{x}=(x_1,x_2,\dots,x_J)^T$ and denote a corresponding projection data (sinogram) by an $I$-dimensional vector $\vec{b}=(b_1,b_2,\dots,b_I)^T$. We denote the system matrix which relates $\vec{x}$ to $\vec{b}$ by $A=\{a_{ij}\}$, where we assume that $I>J$ and $\vec{b}$ is contaminated with statistical noise. Throughout this paper, we assume that the projection data is acquired by the standard parallel-beam geometry so that $\vec{b}$ consists of uniform samples of continuous projection data $p(r,\theta)$. Normally, image reconstruction in this setup can be formulated as the problem of minimizing the Penalized Weighted Least-Squares (P-WLS) cost function expressed as
\begin{eqnarray}
\min_{\vec{x}} f(\vec{x})\equiv\beta \psi(\vec{x})+{{1}\over{2}}\parallel A\vec{x}-\vec{b} \parallel_{W}^2\ \ {\rm subject}\ {\rm to}\ \ \vec{x}\geq 0,
\end{eqnarray}
{\par\noindent}where $W$ denotes $I\times I$ diagonal matrix in which each diagonal element $w_i$ is the inverse of noise variance $\sigma_i^2$, and $\psi(\vec{x})$ is the penalty function to smooth the image [17],[18]. Although there exist several choices in $\psi(\vec{x})$, throughout this paper, we assume that $\psi(\vec{x})$ is the TV penalty function defined by
\begin{eqnarray}
\psi(\vec{x})=\parallel\vec{x}\parallel_{\rm TV}\equiv\sum_{j=1}^J \sqrt{(\vec{h}_j^T\vec{x})^2+(\vec{v}_j^T\vec{x})^2},
\end{eqnarray}
{\par\noindent}where $\vec{h}_j^T\vec{x}$ and $\vec{v}_j^T\vec{x}$ are inner product representations of finite difference operations around the $j$-th pixel along the horizontal and vertical directions, respectively. See Fig. 1 for the detailed definitions of $\vec{h}_j$ and $\vec{v}_j$.
\par Since 2008 when GE Healthcare developed the ASIR (Advanced Statistical Iterative Reconstruction) software, iterative low-dose CT reconstruction has been progressed according to the following three stages.
{\par\noindent}[Image Space Denoising (1-st Generation)] This class of reconstruction methods first perform an FBP reconstruction followed by a smoothing by using an iterative edge-preserving denoising algorithm such as the TV-denoising and the MRF-based denoising.
{\par\noindent}[Iterative FBP Algorithm (2-nd Generation)] Typically, this class of reconstruction methods are based on the following iteration formula
\begin{eqnarray}
\vec{x}^{(k+1)}=P_X[\vec{x}^{(k)}-\gamma(A^TW^{1/2}GW^{1/2}(A\vec{x}^{(k)}-\vec{b})+\beta\nabla\psi(\vec{x}^{(k)}))],
\end{eqnarray}
{\par\noindent}where $P_{X}[\cdot]$ denotes the projection operator onto the positive orthant $\vec{x}\geq 0$ and $G$ is the 1-D ramp filter of FBP reconstruction algorithm having the frequency response $\mid\omega\mid$. Thanks to the introduction of $G$ during the iteration, Eq. (3) converges very fast. However, it is well-known that is cannot exactly minimize the P-WLS cost function of Eq. (1). Another limitation is that the penalty function $\psi(\cdot)$ needs to be differentiable so that the TV penalty, which is non-differentiable, cannot be used.
{\par\noindent}[True Iterative Reconstruction (IR) Algorithm (3-rd Generation)] This class of reconstruction methods try to exactly minimize the P-WLS cost function of Eq. (1). For example, if we use the gradient method, its iteration formula can be expressed as
\begin{eqnarray}
\vec{x}^{(k+1)}=P_X[\vec{x}^{(k)}-\gamma(A^TW(A\vec{x}^{(k)}-\vec{b})+\beta\nabla\psi(\vec{x}^{(k)}))].
\end{eqnarray}
{\par\noindent}The drawback of this method is that its convergence is rather slow compared with the iterative FBP algorithm when no acceleration techniques are incorporated. Another limitation is that the penalty function $\psi(\cdot)$ needs to be differentiable so that the TV penalty cannot be used again.
{\par\noindent}[Preconditioned Gradient Method] An alternative approach which is known in image reconstruction community is to use the iteration formula as
\begin{eqnarray}
\vec{x}^{(k+1)}=P_X[\vec{x}^{(k)}-\gamma M(A^TW(A\vec{x}^{(k)}-\vec{b})+\beta\nabla\psi(\vec{x}^{(k)}))],
\end{eqnarray}
{\par\noindent}where $M$ denotes the 2-D ramp filter in image space having the frequency response $\sqrt{\omega_x^2+\omega_y^2}$ [19]. Unlike Eq. (3), this method converges to an exact minimizer of Eq. (1), but it is rarely used mainly because the 2-D filter is computationally intensive and its design is troublesome.
\par We note that the proposed algorithm derived in the following sections also uses the FBP-type acceleration as in Eq. (3), but it exactly converges to a minimizer of Eq. (1) and also allows to use the TV penalty. Therefore, we believe that the proposed algorithm overcomes the most major drawbacks of existing low-dose CT iterative reconstruction algorithms.
\par In Fig. 2, we show a comparison between the true IR reconstruction and the iterative FBP reconstruction for a numerical chest phantom without a regularization term in the low-dose CT setup. As expected from the theory, it can be observed that the noise property of iterative FBP reconstruction is worse compared with the true IR reconstruction. This result motivates us to develop an alternative FBP-type acceleration technique which is able to exactly minimize the WLS and P-WLS cost functions.
\begin{figure}
\centerline{\includegraphics[height=7cm,clip]{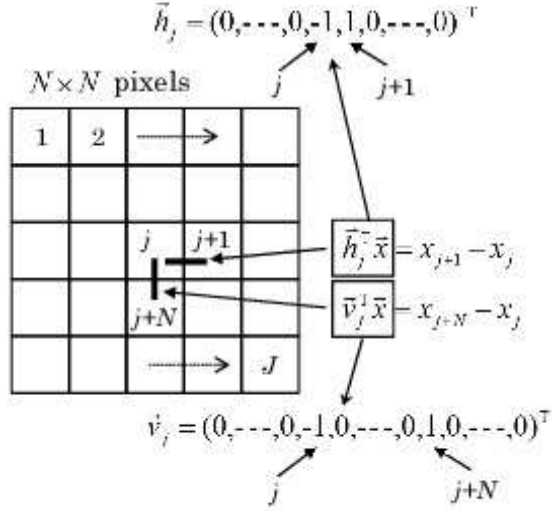}}
\caption{Definitions of the horizontal difference $h_j^T\vec{x}$ and the vertical difference  $v_j^T\vec{x}$ used in the TV penalty function.}
\end{figure}
\begin{figure}
\centerline{\includegraphics[height=10cm,clip]{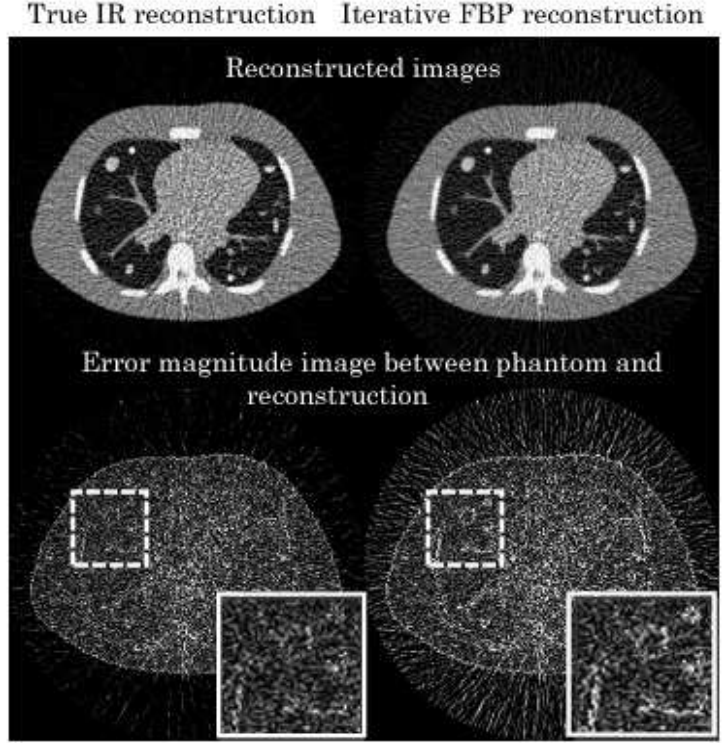}}
\caption{Comparison of reconstructed images and reconstructed error magnitude images between the true IR reconstruction and the iterative FBP reconstruction. The iterative FBP reconstruction suffers from worse noise properties.}
\end{figure}
\subsection{Derivation of Proposed Algorithm}
\par From this section, we derive the proposed algorithm, which is based on reformulating the minimization of Eq. (1) into a saddle point (primal-dual) problem, precondition it, and applying the Alternating Projection Proximal (APP) algorithm which belongs to a class of first-order primal-dual methods [11]. First, Eq. (1) can be reformulated into the following equality constrained minimization by introducing an additional variable $\vec{y}$.
\begin{eqnarray}
\min_{(\vec{x},\vec{y})} (\beta \psi(\vec{x})+{{1}\over{2}}\parallel \vec{y}-\vec{b} \parallel_{W}^2)\ \ {\rm subject}\ {\rm to}\ \ A\vec{x}-\vec{y}=0,\ \vec{x}\geq 0,
\end{eqnarray}
{\par\noindent}where we note that $\vec{y}$ can be interpreted as forward-projected projection data computed from $\vec{x}$. Next, we perform a preconditioning on the constraint $A\vec{x}=\vec{y}$. When we solve Eq. (6) using an iterative algorithm, its convergence speed strongly depends on the condition of linear constraint $A\vec{x}=\vec{y}$. So, we perform the preconditioning by multiplying $A\vec{x}-\vec{y}=0$ by some non-singular $I\times I$ matrix $D^{1/2}$. Then, Eq. (6) can be converted into
\begin{eqnarray}
\min_{(\vec{x},\vec{y})} (\beta \psi(\vec{x})+{{1}\over{2}}\parallel \vec{y}-\vec{b} \parallel_{W}^2)\ \ {\rm subject}\ {\rm to}\ \ D^{1/2}(A\vec{x}-\vec{y})=0,\ \vec{x}\geq 0,
\end{eqnarray}
{\par\noindent}The detailed form of $D^{1/2}$ is not shown here, but we discuss later which choice of $D^{1/2}$ is best to accelerate the convergence. Intuitively, if we choose $D^{1/2}$ in such a way that $D={D}^{T/2}{D}^{1/2}$ approximates the ramp filter in the FBP reconstruction, the final structure of iteration formula resembles the iterative FBP algorithm leading to a fast convergence. Furthermore, we remark that the rewriting from Eq. (6) to Eq. (7) does not alter the problem solution $(\vec{x},\vec{y})$. Next, we reformulate Eq. (7) into a form of saddle point problem [11]-[14]. The procedure is as follows. The Lagrangian function $L(\vec{x},\vec{y},\vec{\mu})$ corresponding to Eq. (7) with respect to the constraint $D^{1/2}(A\vec{x}-\vec{y})=0$ is defined by
\begin{eqnarray}
L(\vec{x},\vec{y},\vec{\mu})=\beta \psi(\vec{x})+{{1}\over{2}}\parallel \vec{y}-\vec{b} \parallel_{W}^2+\vec{\mu}^TD^{1/2}(A\vec{x}-\vec{y}),
\end{eqnarray}
{\par\noindent}where $\vec{\mu}$ is the Lagrange multiplier vector, which is also called the dual variable. Therefore, Eq. (7) can be converted into the saddle point problem\begin{eqnarray}
\max_{\vec{\mu}}\min_{(\vec{x}\geq 0,\vec{y})}L(\vec{x},\vec{y},\vec{\mu})=\beta \psi(\vec{x})+{{1}\over{2}}\parallel \vec{y}-\vec{b} \parallel_{W}^2+\vec{\mu}^TD^{1/2}(A\vec{x}-\vec{y}).
\end{eqnarray}
{\par\noindent}Furthermore, the variable $\vec{y}$ can be eliminated in Eq. (9) by solving the minimization with respect to $\vec{y}$. By performing this computation, we obtain
\begin{eqnarray}
\max_{\vec{\mu}}\min_{\vec{x}\geq 0}L(\vec{x},\vec{\mu})=\beta \psi(\vec{x})-{{1}\over{2}}\parallel D^{T/2}\vec{\mu}\parallel_{W^{-1}}^2-\vec{\mu}^T D^{1/2}\vec{b}+(D^{T/2}\vec{\mu},A\vec{x}),
\end{eqnarray}
{\par\noindent}where $(\cdot,\cdot)$ denotes the inner product. In some literatures, Eq. (10) is called the standard form of saddle point problem [12]. It is also called the primal-dual (PD) problem whereas the original problem of Eq. (1) is called the primal (P) problem [20]-[22]. By using the so-called strong duality theorem in optimization literatures, we can prove the following theorem [20]-[22].
{\par\noindent}[Theorem] Assume that the penalty function $\psi(\vec{x})$ is a possibly non-differentiable convex function. We denote the solution to the primal problem (Eq. (1)) by $\vec{x}^{({\rm P})}$. We denote the solution to the primal-dual problem (Eq. (10)) by $(\vec{x}^{({\rm PD})},\vec{\mu}^{({\rm PD})})$. Then, the following two properties hold.
{\par\noindent}(1) $\vec{x}^{({\rm P})}=\vec{x}^{({\rm PD})}$
{\par\noindent}(2) $L(\vec{x}^{({\rm PD})},\vec{\mu}^{(\rm{PD})})=f(\vec{x}^{({\rm P})})$
{\par\noindent}From this theorem, if we succeed in exactly solving the saddle point problem of Eq. (10), its solution vector $\vec{x}^{({\rm PD})}$ is also a solution of the original problem of Eq. (1), {\it i.e.} $\vec{x}^{({\rm P})}$. In summary of the above discussion, we need to solve the following saddle point problem.
\begin{eqnarray}
& &\ \ \ \ \max_{\vec{\mu}}\min_{\vec{x}} L(\vec{x},\vec{\mu})=g(\vec{x})-h(\vec{\mu})+(D^{T/2}\vec{\mu},A\vec{x}), \nonumber \\
& &g(\vec{x})=\beta\psi(\vec{x})+i(\vec{x}),\ h(\vec{\mu})={{1}\over{2}}\parallel D^{T/2}\vec{\mu}\parallel_{W^{-1}}^2+\vec{\mu}^T D^{1/2}\vec{b},
\end{eqnarray}
{\par\noindent}where the non-negativity constraint $\vec{x}\geq 0$ was put into the cost function $g(\vec{x})$ as the indicator function $i(\vec{x})$ defined by
\begin{eqnarray}
i(\vec{x})=\cases{0 & (if $\vec{x}\geq 0$) \cr \infty & (otherwise)}.
\end{eqnarray}
\par The proposed algorithm is based on solving Eq. (11) instead of Eq. (1). There exist a variety of iterative methods to solve Eq. (11) in optimization literatures such as the classical multiplier method, augmented Lagrangian method, Alternating Direction Method of Multiplier (ADMM) method, etc [20]-[23]. Among them, we use a class of iterative methods called the first-order primal-dual methods, because they require relatively simple computations per iteration [11]-[14]. The basic overall structure of this iterative method is to repeat an update of primal variable $\vec{x}$ along the descent direction of Lagrangian $-\partial L(\vec{x},\vec{\mu})/\partial\vec{x}$, {\it i.e.} the descent direction of $g(\vec{x})+(D^{T/2}\vec{\mu},A\vec{x})$, and an update of dual variable $\vec{\mu}$ along the ascent direction of Lagrangian $\partial L(\vec{x},\vec{\mu})/\partial\vec{\mu}$, {\it i.e.} the ascent direction of $-h(\vec{\mu})+(D^{T/2}\vec{\mu},A\vec{x})$, alternately with some additional extrapolation step. The role of extrapolation step is to guarantee that each update from $(\vec{x}^{(k)},\vec{\mu}^{(k)})$ to $(\vec{x}^{(k+1)},\vec{\mu}^{(k+1)})$ becomes a non-expansive mapping leading to a convergence to the saddle point. In each update along the descent or ascent direction, either the gradient step or the proximal step can be used dependent on differentiabilities of $g(\vec{x})$ and $h(\vec{\mu})$. Typically, there exist four variations in the first-order primal-dual methods, which are the Chambolle-Pock algorithm, Generalized Iterative Soft-Thresholding (GIST) algorithm, Alternating Projection Proximal (APP) algorithm, and Alternating Extragradient (AE) algorithm [11]-[14]. The differences in used operations among the four algorithms are briefly summarized in Table 1. 
\begin{table}
\begin{center}
\begin{tabular}{|p{12em}|p{14em}|p{4em}|p{4em}|p{6em}|}
\hline
Name & Required differentiability & Primal update & Dual update & Extrapolation \\
\hline\hline
Alternating Projection Proximal (APP) [11] & $g(\vec{x})$ can be non-differentiable, $h(\vec{\mu})$ needs to be differentiable & proximal & gradient & dual space \\ \hline
Chambolle-Pock (CP) [12] & both $g(\vec{x})$ and $h(\vec{\mu})$ can be non-differentiable & proximal & proximal & primal space \\ \hline
Generalized Iterative Soft-Thresholding (GIST) [13] & $g(\vec{x})$ needs to be differentiable, $h(\vec{\mu})$ can be non-differentiable & gradient & proximal & dual space \\ \hline
Alternating Extragradient (AE) [14] & both $g(\vec{x})$ and $h(\vec{\mu})$ need to be differentiable & gradient & gradient & dual space \\ \hline
\end{tabular}
\end{center}
\caption{Differences among the four first-order primal-dual algorithms.}
\end{table}
{\par\noindent}In our problem, as is clear from Eq. (11), $g(\vec{x})$ is non-differentiable because it corresponds to the TV penalty (combined with the non-negativity constraint) whereas $h(\vec{\mu})$ is differentiable (quadratic form). Thus, we decided to employ the APP algorithm [11]. For general (non-differentiable) $g(\vec{x})$ and (differentiable) $h(\vec{\mu})$, the iteration formula of APP algorithm in its original form is summarized as follows.
\begin{eqnarray}
& &{\rm [Extrapolation\ Step]} \nonumber \\
& & \ \ \ \ \ \ \ \ \overline{\vec{\mu}}^{(k+1)}=\vec{\mu}^{(k)}+\sigma(D^{1/2}A\vec{x}^{(k)}-\nabla h(\vec{\mu}^{(k)})) \\
& & {\rm [Primal\ Update\ (Proximal)]} \nonumber \\
& & \ \ \ \ \ \ \ \ \vec{x}^{(k+1)}={\rm prox}_{\tau g}(\vec{x}^{(k)}-\tau A^TD^{T/2}\overline{\vec{\mu}}^{(k+1)})\\
& & {\rm [Dual\ Update\ (Gradient)]} \nonumber \\
& & \ \ \ \ \ \ \ \ \vec{\mu}^{(k+1)}=\vec{\mu}^{(k)}+\sigma(D^{1/2}A\vec{x}^{(k+1)}-\nabla h(\vec{\mu}^{(k)})),
\end{eqnarray}
{\par\noindent}where $k$ is the iteration number, and $\tau>0$, $\sigma>0$ are stepsize parameters corresponding to the primal update and the dual update, respectively. To guarantee the convergence, we assume that $\tau$ and $\sigma$ are selected such that
\begin{eqnarray}
0<\sigma<{{2}\over{L}},\ 0<\tau<{{1}\over{\sigma\parallel D^{1/2}AA^TD^{T/2}\parallel}},
\end{eqnarray}
{\par\noindent}where $L$ is Lipschitz constant of the gradient $\nabla h(\vec{\mu})$ [11]. Furthermore, the operator ${\rm prox}_{\tau g}(\cdot)$ appearing in Eq. (14) denotes the proximity (prox) operator defined by
\begin{eqnarray}
\vec{x}={\rm prox}_{\tau g}(\vec{z})\equiv{\rm arg}\min_{\vec{x}}(\tau g(\vec{x})+{{1}\over{2}}\parallel\vec{x}-\vec{z}\parallel^2)={\rm arg}\min_{\vec{x}\geq 0}(\tau\beta\psi(\vec{x})+{{1}\over{2}}\parallel\vec{x}-\vec{z}\parallel^2),
\end{eqnarray}
{\par\noindent}where $\tau$ is called the stepsize parameter [24]. We will describe later on how to compute the prox operator when $\psi(\vec{x})$ is the TV penalty. We note that Eq. (13) represents the extrapolation step to mathematically guarantee the convergence, Eq. (14) is the primal update using the prox operator, and Eq. (15) is the dual update using the gradient step. The iteration formula in our special case of Eq. (11) is obtained by applying the general algorithm of Eqs. (13)-(15) followed by the variable change $\vec{\mu}\leftarrow D^{T/2}\vec{\mu}$ and rewriting the extrapolation step into a much simpler form. The resulting iteration formula is expressed as
\begin{eqnarray}
& &{\rm [Extrapolation\ Step]} \nonumber \\
& & \ \ \ \ \ \ \ \ \overline{\vec{\mu}}^{(k+1)}=\vec{\mu}^{(k)}+\sigma D(A\vec{x}^{(k)}-\vec{b}-W^{-1}\vec{\mu}^{(k)})=2\vec{\mu}^{(k)}-\vec{\mu}^{(k-1)}-\sigma DW^{-1}(\vec{\mu}^{(k)}-\vec{\mu}^{(k-1)}) \\
& & {\rm [Primal\ Update]} \nonumber \\
& & \ \ \ \ \ \ \ \ \vec{x}^{(k+1)}={\rm prox}_{\tau g}(\vec{x}^{(k)}-\tau A^T\overline{\vec{\mu}}^{(k+1)})\\
& & {\rm [Dual\ Update]} \nonumber \\
& & \ \ \ \ \ \ \ \ \vec{\mu}^{(k+1)}=\vec{\mu}^{(k)}+\sigma D(A\vec{x}^{(k+1)}-\vec{b}-W^{-1}\vec{\mu}^{(k)}),
\end{eqnarray}
{\par\noindent}where $D=D^{T/2}D^{1/2}$ and we simplified the extrapolation step using the fact that the $\sigma D(A\vec{x}^{(k)}-\vec{b})$ term in the extrapolation step is same as $\sigma D(A\vec{x}^{(k+1)}-\vec{b})$ appearing in the previous dual update. The algorithm derivation was finished. With respect to computational requirements, the dominant computations in implementing Eqs. (18)-(20) are a forward projection $A$, a backprojection $A^T$, and a prox operator associated with the TV penalty in each iteration. Therefore, we expect that the computational costs of the proposed algorithm are similar to those of the other iterative reconstruction algorithms which use the TV penalty.
\par In implementing Eqs. (18)-(20), there still exist two unclear points. The first point is how to compute ${\rm prox}_{\tau g}(\cdot)$ in the primal update, which amounts to solving the minimization of Eq. (17). When $\psi(\vec{x})$ is the TV penalty of Eq. (2), Eq. (17) is same as the so-called TV denoising problem (ROF problem) investigated in [25],[26], to which a variety of efficient algorithms are available. In our implementations, we use Chambolle's projection algorithm for this purpose modified in such a way that the non-negativity constraint $\vec{x}\geq 0$ is incorporated [26]. The second unclear point concerns how to design the preconditioning matrix $D$ to achieve a fast convergence. Intuitively, from Eqs. (18)-(20), it can be inspired that a nice choice of $D$ is the ramp filter in the FBP reconstruction, because, in this case, the first iteration initialized with $\vec{x}^{(0)}=0,\vec{\mu}^{(0)}=0$ coincides with the FBP reconstruction followed by the TV denoising. However, this is not the best choice when the projection data $\vec{b}$ contains statistical noise. We discuss this issue below. First, we note that $D$ needs to be a positive definite matrix, because $D$ is required to be expressed as $D=D^{T/2}D^{1/2}$. By forgetting the extrapolation step (for some moment) and combining Eqs. (19) and (20), the approximate iteration formula with respect to the dual variable $\vec{\mu}$ is expressed as
\begin{eqnarray}
\vec{\mu}^{(k+1)}\approx\vec{\mu}^{(k)}+\sigma D(A{\rm prox}_{\tau g}(\vec{x}^{(k)}-\tau A^T\vec{\mu}^{(k)})-\vec{b}-W^{-1}\vec{\mu}^{(k)}).
\end{eqnarray}
{\par\noindent}Furthermore, neglecting the prox operator, we have
\begin{eqnarray}
\vec{\mu}^{(k+1)}&\approx&\vec{\mu}^{(k)}+\sigma D(A(\vec{x}^{(k)}-\tau A^T\vec{\mu}^{(k)})-\vec{b}-W^{-1}\vec{\mu}^{(k)}) \nonumber \\
&\approx&\vec{\mu}^{(k)}-\sigma D((\tau A A^T+W^{-1})\vec{\mu}^{(k)}-(A\vec{x}^{(k)}-\vec{b})).
\end{eqnarray}
{\par\noindent}Equation (22) implies that the meaning of $\vec{\mu}$ is the noise compoment in the projection data and the APP method is implicitly solving the equation $(\tau AA^T+W^{-1})\vec{\mu}=A\vec{x}^{(k)}-\vec{b}$ at each iteration $k$ by using the preconditioned iteration of Eq. (22). From this observation, the best choice of $D$ is clearly given by
\begin{eqnarray}
D=(\tau A A^T+W^{-1})^{-1}.
\end{eqnarray}
{\par\noindent}However, Eq. (23) is not practical because its computation requires a large matrix inverse. So, we use a shift-invariant approximation $W^{-1}\approx\kappa I$ where $\kappa $ is the average value of diagonal elements of $W^{-1}$. Then, noting that $AA^T$ is the blurring operator $2m/\mid\omega\mid$ in projection data space ($m$ is the number of view angles over $0\leq\theta\leq 180^{\circ}$ in the parallel-beam projection data), $D$ becomes the frequency domain filter expressed as
\begin{eqnarray}
& &D=(\tau A A^T+\kappa I)^{-1}=F^{-1}H(\omega)F \nonumber \\
& &H(\omega)={{\mid \omega\mid}\over{2m\tau+\kappa\mid\omega\mid}},
\end{eqnarray}
{\par\noindent}where $F$ and $F^{-1}$ denote the 1-D Fourier transform with respect to the radial variable in the projection data space and its inverse, respectively. We note that Eq. (24) is the smoothed ramp filter. We show the frequency response $H(\omega)$ with varying parameter $\tau$ in Fig. 3. Finally, we summarize that the smoothed ramp filter is the best preconditioning matrix $D$ in the case of low-dose CT reconstruction. We experimentally confirmed that there exists an essential difference between the ordinary (non-smoothed) ramp filter and the smoothed ramp filter in the convergence speed of cost function value $f(\vec{x})$.
\begin{figure}
\centerline{\includegraphics[height=5cm,clip]{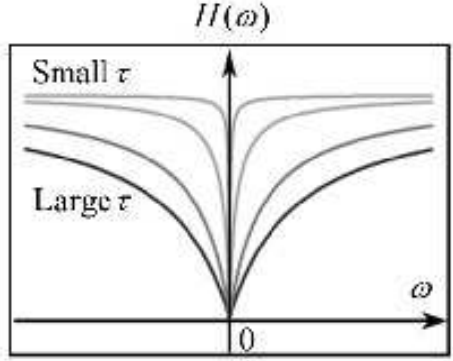}}
\caption{Frequency response of the preconditioning filter $H(\omega)$ used in the proposed algorithm for the low-dose CT.}
\end{figure}
\subsection{Summary of Proposed Algorithm}
\par We summarize the proposed algorithm for the low-dose CT in Algorithm 1. As a special case of $\beta=0$, {\it i.e.} absence of the TV penalty term, we obtain a fast FBP-preconditioned algorithm to exactly minimize the WLS cost function, which is also summarized in Algorithm 1'.
\begin{table}
\begin{center}
\begin{tabular}{p{40em}}
\hline
{\par\noindent}Algorithm 1: Low-Dose CT TV-Regularized WLS Reconstruction \\
\hline\hline
{\par\noindent}Prepare the preconditioning matrix $D$. Give initial primal and dual vectors by $\vec{x}^{(0)}=0$ and $\vec{\mu}^{(0)}=0$. Set stepsize parameters $(\sigma>0,\tau>0)$ such that $0<\sigma<2/\parallel D^{1/2}W^{-1}D^{T/2}\parallel$ and $0<\tau<1/(\sigma\parallel D^{1/2}AA^T D^{T/2}\parallel)$. Execute the following steps for $k=0,1,2,\dots$.
{\par\noindent}[Step 1] Extrapolation Step \\
{\par\noindent}$\ \ \ \ \ \overline{\vec{\mu}}^{(k+1)}=\cases{-\sigma D\vec{b} & (if $k=0$) \cr 2\vec{\mu}^{(k)}-\vec{\mu}^{(k-1)}-\sigma DW^{-1}(\vec{\mu}^{(k)}-\vec{\mu}^{(k-1)}) & (if $k\neq 0$)}$ \\
{\par\noindent}[Step 2] Primal Update \\
{\par\noindent}$\ \ \ \ \ \vec{x}^{(k+1)}={\rm prox}_{\tau g}(\vec{x}^{(k)}-\tau A^T\overline{\vec{\mu}}^{(k+1)})$ \\
{\par\noindent}$\displaystyle\ \ \ \ \ \ \ \ \ \ \ \ \ \ ={\rm arg}\min_{\vec{x}\geq 0}(\tau\beta\psi({\vec{x}})+{{1}\over{2}}\parallel\vec{x}-(\vec{x}^{(k)}-\tau A^T\overline{\vec{\mu}}^{(k+1)})\parallel^2)$ \\
{\par\noindent}[Step 3] Dual Update \\
{\par\noindent}$\ \ \ \ \ \vec{\mu}^{(k+1)}=\vec{\mu}^{(k)}+\sigma D(A\vec{x}^{(k+1)}-\vec{b}-W^{-1}\vec{\mu}^{(k)})$ \\
\hline
\end{tabular}
\end{center}
\end{table}
\begin{table}
\begin{center}
\begin{tabular}{p{40em}}
\hline
{\par\noindent}Algorithm 1': Low-Dose CT WLS Reconstruction \\
\hline\hline
{\par\noindent}Prepare the preconditioning matrix $D$. Give initial primal and dual vectors by $\vec{x}^{(0)}=0$ and $\vec{\mu}^{(0)}=0$. Set stepsize parameters $(\sigma>0,\tau>0)$ such that $0<\sigma<2/\parallel D^{1/2}W^{-1}D^{T/2}\parallel$ and $0<\tau<1/(\sigma\parallel D^{1/2}AA^T D^{T/2}\parallel)$. Execute the following steps for $k=0,1,2,\dots$.
{\par\noindent}[Step 1] Extrapolation Step \\
{\par\noindent}$\ \ \ \ \ \overline{\vec{\mu}}^{(k+1)}=\cases{-\sigma D\vec{b} & (if $k=0$) \cr 2\vec{\mu}^{(k)}-\vec{\mu}^{(k-1)}-\sigma DW^{-1}(\vec{\mu}^{(k)}-\vec{\mu}^{(k-1)}) & (if $k\neq 0$)}$ \\
{\par\noindent}[Step 2] Primal Update \\
{\par\noindent}$\ \ \ \ \ \vec{x}^{(k+1)}=P_{X}[\vec{x}^{(k)}-\tau A^T\overline{\vec{\mu}}^{(k+1)}]$ \\
{\par\noindent}[Step 3] Dual Update \\
{\par\noindent}$\ \ \ \ \ \vec{\mu}^{(k+1)}=\vec{\mu}^{(k)}+\sigma D(A\vec{x}^{(k+1)}-\vec{b}-W^{-1}\vec{\mu}^{(k)})$ \\
\hline
\end{tabular}
\end{center}
\end{table}

\section{PROPOSED ALGORITHM FOR FEW-VIEW CT RECONSTRUCTION}
\par The algorithm derivation used in Section 2 based on the first-order primal-dual methods combined with the preconditioning is a rather general framework so that it can be applied to develop a fast convergent reconstruction algorithm in the few-view CT formulation. We denote an image by a $J$-dimensional vector $\vec{x}=(x_1,x_2,\dots,x_J)^T$ and denote a corresponding projection data (sinogram) by an $I$-dimensional vector $\vec{b}=(b_1,b_2,\dots,b_I)^T$. We denote the system matrix which relates $\vec{x}$ to $\vec{b}$ by $A=\{a_{ij}\}$, where we assume that $I<J$ and $\vec{b}$ does not contain statistical noise. Normally, image reconstruction in this setup can be formulated as the problem of minimizing a penalty function $\psi(\vec{x})$ under the linear constraint $A\vec{x}=\vec{b}$ as
\begin{eqnarray}
\min_{\vec{x}} \psi(\vec{x})\ \ {\rm subject}\ {\rm to}\ A\vec{x}=\vec{b},\ \vec{x}\geq 0,
\end{eqnarray}
{\par\noindent}where we assume that $\psi(\cdot)$ is the TV penalty defined by Eq. (2) throughout this section. Very often, in engineering literatures, the problem of Eq. (25) is converted into the unconstrained minimization $\min_{\vec{x}\geq 0}\ (\beta\psi(\vec{x})+\parallel A\vec{x}-\vec{b}\parallel^2)$ followed by solving it using a technique of unconstrained optimizations. However, the solution obtained in this way can be quite different from that of the original constrained problem of Eq. (25). The proposed algorithm below allows us to exactly solve Eq. (25).
\par We derive the proposed algorithm to solve Eq. (25) using the same mathematical framework as in Section 2 below. First, we introduce the $I\times I$ non-singular preconditioning matrix $D^{1/2}$ into Eq. (25) as
\begin{eqnarray}
\min_{\vec{x}} \psi(\vec{x})\ \ {\rm subject}\ {\rm to}\ \ D^{1/2}(A\vec{x}-\vec{b})=0,\ \vec{x}\geq 0.
\end{eqnarray}
{\par\noindent}We note that the solution to Eq. (25) is same as the solution to Eq. (26). The Lagrangian function $L(\vec{x},\vec{\mu})$ corresponding to Eq. (26) with respect to the constraint $D^{1/2}(A\vec{x}-\vec{b})=0$ is defined by
\begin{eqnarray}
L(\vec{x},\vec{\mu})=\psi(\vec{x})+\vec{\mu}^TD^{1/2}(A\vec{x}-\vec{b}),
\end{eqnarray}
{\par\noindent}where $\vec{\mu}$ is the Lagrange multiplier vector (the dual variable). Therefore, Eq. (26) can be converted into the saddle point problem
\begin{eqnarray}
\max_{\vec{\mu}}\min_{\vec{x}}L(\vec{x},\vec{\mu})=g(\vec{x})-h(\vec{\mu})+(D^{T/2}\vec{\mu},A\vec{x}),\ g(\vec{x})=\psi(\vec{x})+i(\vec{x}),\ h(\vec{\mu})=\vec{\mu}^TD^{1/2}\vec{b},
\end{eqnarray}
{\par\noindent}where $i(\vec{x})$ is the indicator function defined by Eq. (12). Since the structure of Eq. (28) is same as that of Eq. (11) (there exists only a difference in the form of $h(\vec{\mu})$), we can use the same approach as in the low-dose CT case to develop an iterative algorithm. Applying the APP algorithm given by Eqs. (13)-(15) to Eq. (28) yields the following iterative algorithm.
\begin{eqnarray}
& &{\rm [Extrapolation\ Step]} \nonumber \\
& & \ \ \ \ \ \ \ \ \overline{\vec{\mu}}^{(k+1)}=\vec{\mu}^{(k)}+\sigma D(A\vec{x}^{(k)}-\vec{b})=2\vec{\mu}^{(k)}-\vec{\mu}^{(k-1)} \\
& & {\rm [Primal\ Update]} \nonumber \\
& & \ \ \ \ \ \ \ \ \vec{x}^{(k+1)}={\rm prox}_{\tau g}(\vec{x}^{(k)}-\tau A^T\overline{\vec{\mu}}^{(k+1)})\\
& & {\rm [Dual\ Update]} \nonumber \\
& & \ \ \ \ \ \ \ \ \vec{\mu}^{(k+1)}=\vec{\mu}^{(k)}+\sigma D(A\vec{x}^{(k+1)}-\vec{b}).
\end{eqnarray}
{\par\noindent}The algorithm derivation was finished. We note that the only difference between the low-dose CT case and the few-view CT case lies in the absence of $W^{-1}\vec{\mu}^{(k)}$ term in the latter from the comparison between Eqs. (18)-(20) and Eqs. (29)-(31). Next, we discuss about the preconditioning matrix $D$ to achieve a fast convergence. By following the same discussion as in the low-dose CT case, thanks to the absence of $W^{-1}\vec{\mu}^{(k)}$ term, the best preconditioning matrix $D$ in Eqs. (29)-(31) can be shown to be the ramp filter in the FBP reconstruction, which is expressed as
\begin{eqnarray}
& &D=(\tau A A^T)^{-1}=F^{-1}H(\omega)F \nonumber \\
& &H(\omega)={{\mid \omega\mid}\over{2m\tau}},
\end{eqnarray}
{\par\noindent}where we note again that $m$ is the number of view angles over $0\leq\theta\leq 180^{\circ}$ in the parallel-beam projection data, and $F$ and $F^{-1}$ denote the 1-D Fourier transform with respect to the radial variable in the projection data space and its inverse, respectively. With this choice of $D$, the first iteration initialized with $\vec{x}^{(0)}=0,\vec{\mu}^{(0)}=0$ coincides with the FBP reconstruction followed by the TV smoothing (denoising).
\subsection{Summary of Proposed Algorithm}
\par We summarize the proposed algorithm for the few-view CT in Algorithm 2.
\begin{table}
\begin{center}
\begin{tabular}{p{40em}}
\hline
{\par\noindent}Algorithm 2: Few-View CT TV-Regularized Reconstruction \\
\hline\hline
{\par\noindent}Prepare the preconditioning matrix $D$. Give initial primal and dual vectors by $\vec{x}^{(0)}=0$ and $\vec{\mu}^{(0)}=0$. Set stepsize parameters $(\sigma>0,\tau>0)$ such that $0<\sigma\tau<1/\parallel D^{1/2}AA^TD^{T/2}\parallel$. Execute the following steps for $k=0,1,2,\dots$.
{\par\noindent}[Step 1] Extrapolation Step \\
{\par\noindent}$\ \ \ \ \ \overline{\vec{\mu}}^{(k+1)}=\cases{-\sigma D\vec{b} & (if $k=0$) \cr 2\vec{\mu}^{(k)}-\vec{\mu}^{(k-1)} & (if $k\neq 0$)}$\\
{\par\noindent}[Step 2] Primal Update \\
{\par\noindent}$\ \ \ \ \ \vec{x}^{(k+1)}={\rm prox}_{\tau g}(\vec{x}^{(k)}-\tau A^T\overline{\vec{\mu}}^{(k+1)})$ \\
{\par\noindent}$\displaystyle\ \ \ \ \ \ \ \ \ \ \ \ \ \ ={\rm arg}\min_{\vec{x}\geq 0}(\tau\psi({\vec{x}})+{{1}\over{2}}\parallel\vec{x}-(\vec{x}^{(k)}-\tau A^T\overline{\vec{\mu}}^{(k+1)})\parallel^2)$ \\
{\par\noindent}[Step 3] Dual Update \\
{\par\noindent}$\ \ \ \ \ \vec{\mu}^{(k+1)}=\vec{\mu}^{(k)}+\sigma D(A\vec{x}^{(k+1)}-\vec{b})$ \\
\hline
\end{tabular}
\end{center}
\end{table}
\section{SIMULATION STUDIES}
\par We have performed two simulation studies to demonstrate performances of Algorithm 1 in the low-dose CT reconstruction and Algorithm 2 in the few-view CT reconstruction. The details of simulation studies are summarized below.

{\par\noindent}[Low-Dose CT Simulation] We used a numerical phantom called the spot phantom consisting of $256\times 256$ (pixels) and a single slice of chest-scan CT image consisting of 320$\times$320 (pixels). The simulated projection data was computed with 1,200 (angles) by the parallel-beam geometry followed by adding noise which follows transmission Poisson statistics, from which image reconstructions were performed. We compared the proposed algorithm with the following three competitive algorithms.
{\par\noindent}(a) Gradient descent algorithm: The standard gradient descent algorithm to minimize the TV-regularized WLS cost function of Eq. (1) was implemented. To deal with the non-differentiability of TV penalty, the TV norm was approximated by the standard smoothing technique as
\begin{eqnarray}
\psi(\vec{x})\approx\sum_{j=1}^J(\sqrt{(\vec{h}_j^T\vec{x})^2+(\vec{v}_j^T\vec{x})^2+\epsilon}-\sqrt{\epsilon}),
\end{eqnarray}
{\par\noindent}where $\epsilon>0$ is a small positive number to control the degree of smoothing.
{\par\noindent}(b) Iterative FBP algorithm [15],[16]: The iterative FBP algorithm was implemented mainly to demonstrate that it cannot accurately minimize the TV-regularized WLS cost function of Eq. (1) and there exists an image degradation although it converges very fast.
{\par\noindent}(c) Generalized Iterative Soft-Thresholding (GIST) algorithm [13]: We also implemented one of newest iterative algorithms developed by applied mathematicians, which is able to exactly minimize the TV-regularized WLS cost function of Eq. (1). Since this algorithm is of simultaneous update type such as the SIRT and SART algorithms, we expect that its convergence is rather slow.
{\par\noindent}[Few-View CT Simulation] We used the spot phantom same as in the low-dose CT simulation. The simulated projection data was computed with 32 (angles) by the parallel-beam geometry, from which image reconstructions were performed. We compared the proposed algorithm with the following two competitive algorithms.
{\par\noindent}(a) Chambolle-Pock algorithm [4],[12]: We implemented one of newest iterative algorithms developed by applied mathematicians, which is able to exactly solve the minimization of TV penalty under the constraints $A\vec{x}=\vec{b},\ \vec{x}\geq 0$. Since this algorithm is of simultaneous update type as in the GIST algorithm, we expect that its convergence is rather slow.
{\par\noindent}(b) ART algorithm [27]: We also implemented the standard ART algorithm without the TV penalty, which is known to converge fast thanks to its row-action (sequential update) structure.
\par In Figs. 4-5, we show reconstructed images in the low-dose CT case together with corresponding Root Mean Squares (RMSE) reconstruction errors defined by
\begin{eqnarray}
{\rm RMSE}=\sqrt{{{\displaystyle\sum_{j=1}^J (x_j-x^{\rm (true)}_j)^2}\over{J}}},
\end{eqnarray}
{\par\noindent}where $x_j$ denotes the reconstructed pixel value and $x^{({\rm true})}_j$ denotes the true pixel value. Also, in Fig. 6, we show the corresponding convergence properties of each reconstruction algorithm. These figures clearly demonstrate that Algorithm 1 converges to the exact minimizer of the TV-regularized WLS cost function very fast. In particular, it can be observed that the iterative FBP algorithm converges to an image with worse RMSE value although its convergence is very fast, whereas Algorithm 1 converges to the exact minimizer with a much faster speed compared with the other slow algorithms such as the gradient descent and GIST algorithms thanks to the FBP-type preconditioning.
\par In Fig. 7, we show reconstructed images in the few-view CT case together with corresponding RMSE reconstruction errors. Also, in Fig. 8, we show the corresponding convergence properties of each reconstruction algorithm. These figures clearly demonstrate that Algorithm 2 converges to an image with much smaller RMSE value compared with the ART algorithm, and its convergence is much faster than the Chambolle-Pock algorithm again thanks to the FBP-type preconditioning.
\begin{figure}
\centerline{\includegraphics[height=14cm,clip]{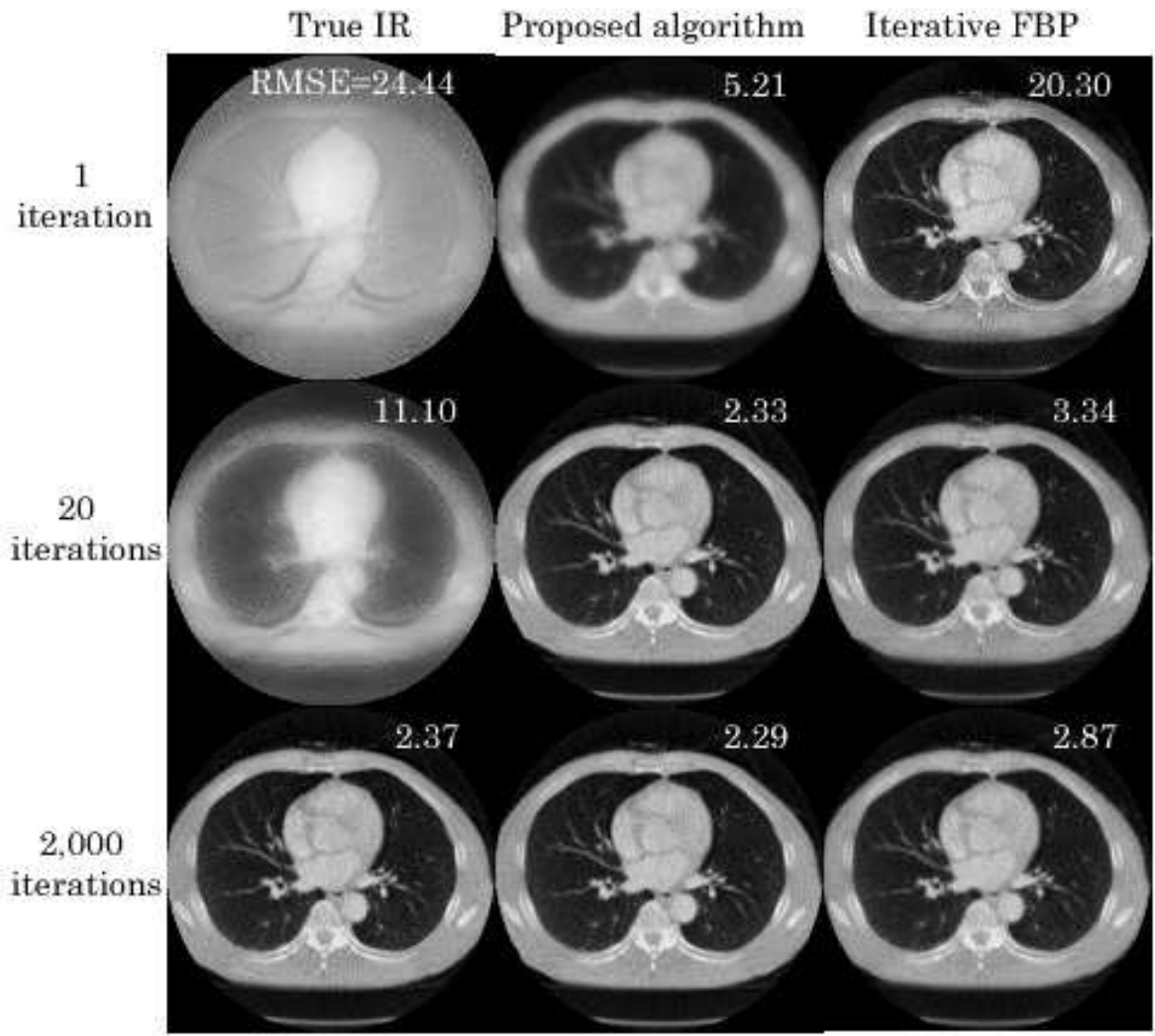}}
\caption{Reconstructed images of chest CT phantom in the low-dose CT experiment (1,200 projection data). Algorithm 1 was compared with the gradient descent algorithm and the iterative FBP algorithm. The numbers in images represent RMSE reconstruction errors.}
\end{figure}
\begin{figure}
\centerline{\includegraphics[height=9cm,clip]{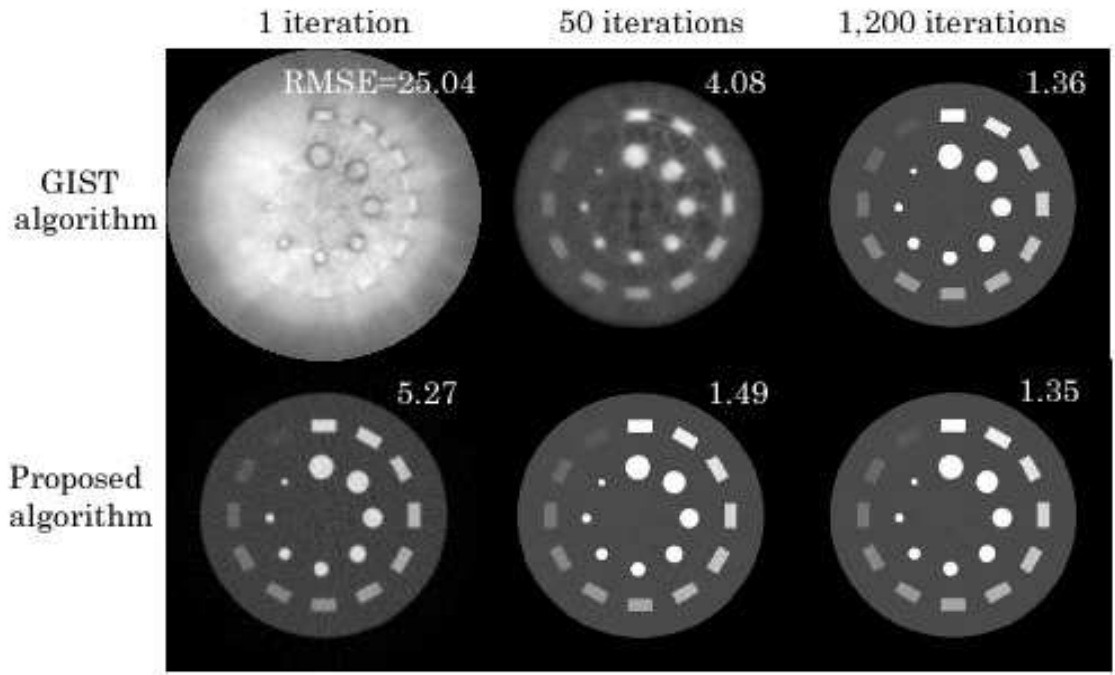}}
\caption{Reconstructed images of spot phantom in the low-dose CT experiment. Algorithm 1 was compared with the GIST algorithm. The numbers in images represent RMSE reconstruction errors.}
\end{figure}
\begin{figure}
\centerline{\includegraphics[height=15cm,clip]{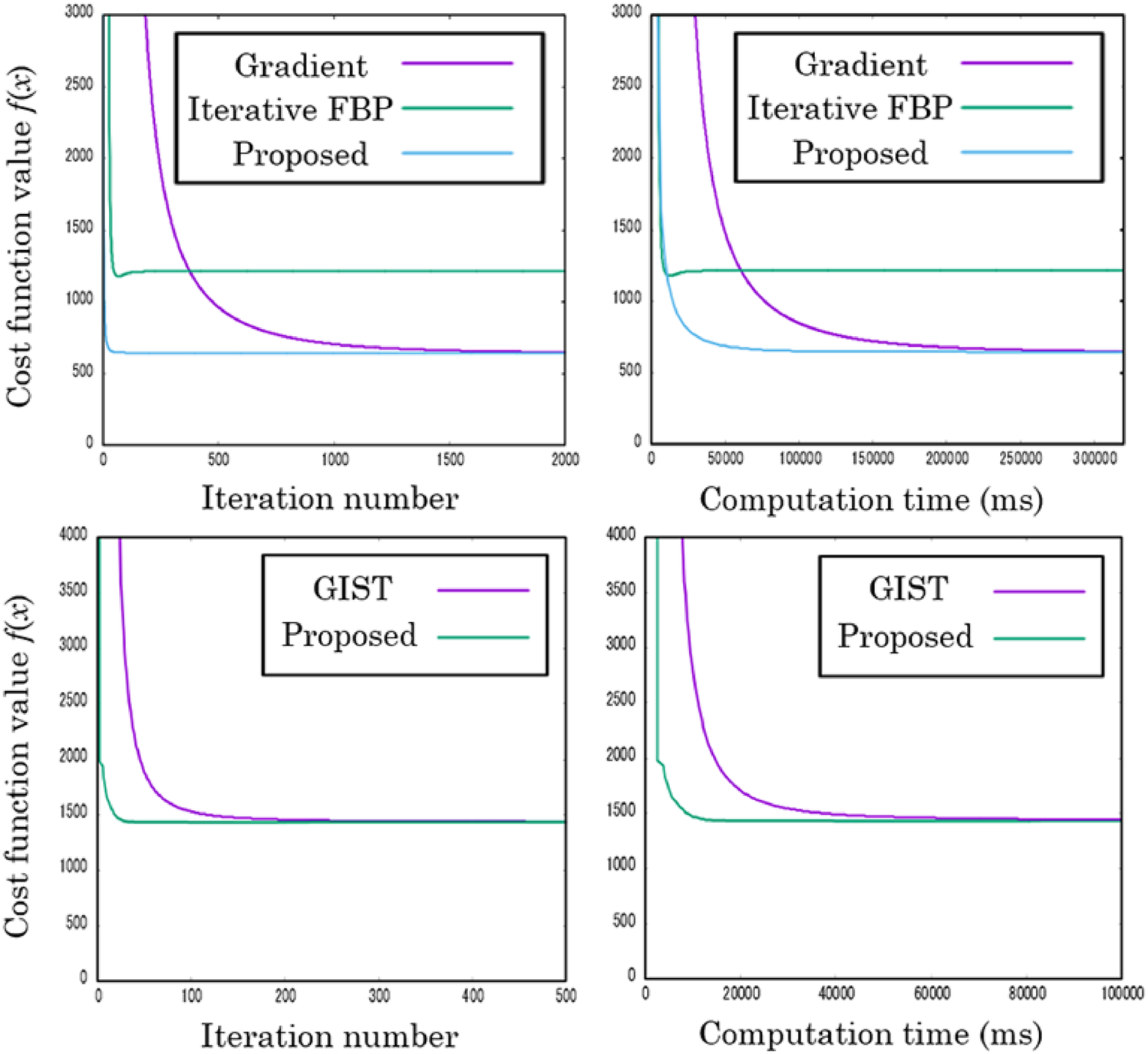}}
\caption{Comparison of convergence properties among Algorithm 1, gradient descent algorithm, iterative FBP algorithm, and the GIST algorithm in the low-dose CT experiment. Iteration number (computational time) versus cost function value $f(\vec{x})$ was plotted.}
\end{figure}
\begin{figure}
\centerline{\includegraphics[height=14cm,clip]{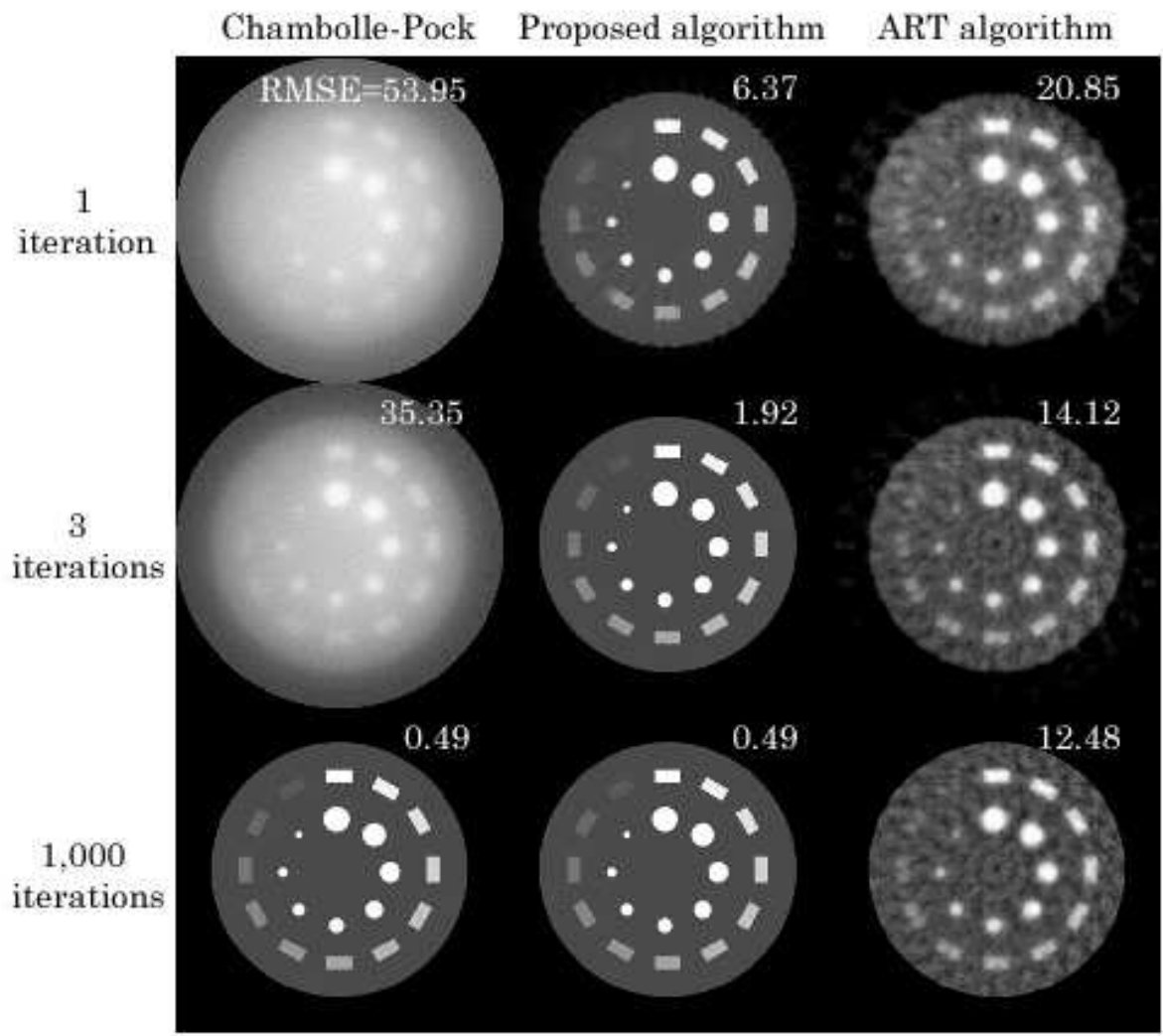}}
\caption{Reconstructed images of spot phantom in the few-view CT experiment (32 projection data). Algorithm 2 was compared with the Chambolle-Pock algorithm and the ART algorithm. The numbers in images represent RMSE reconstruction errors.}
\end{figure}
\begin{figure}
\centerline{\includegraphics[height=7cm,clip]{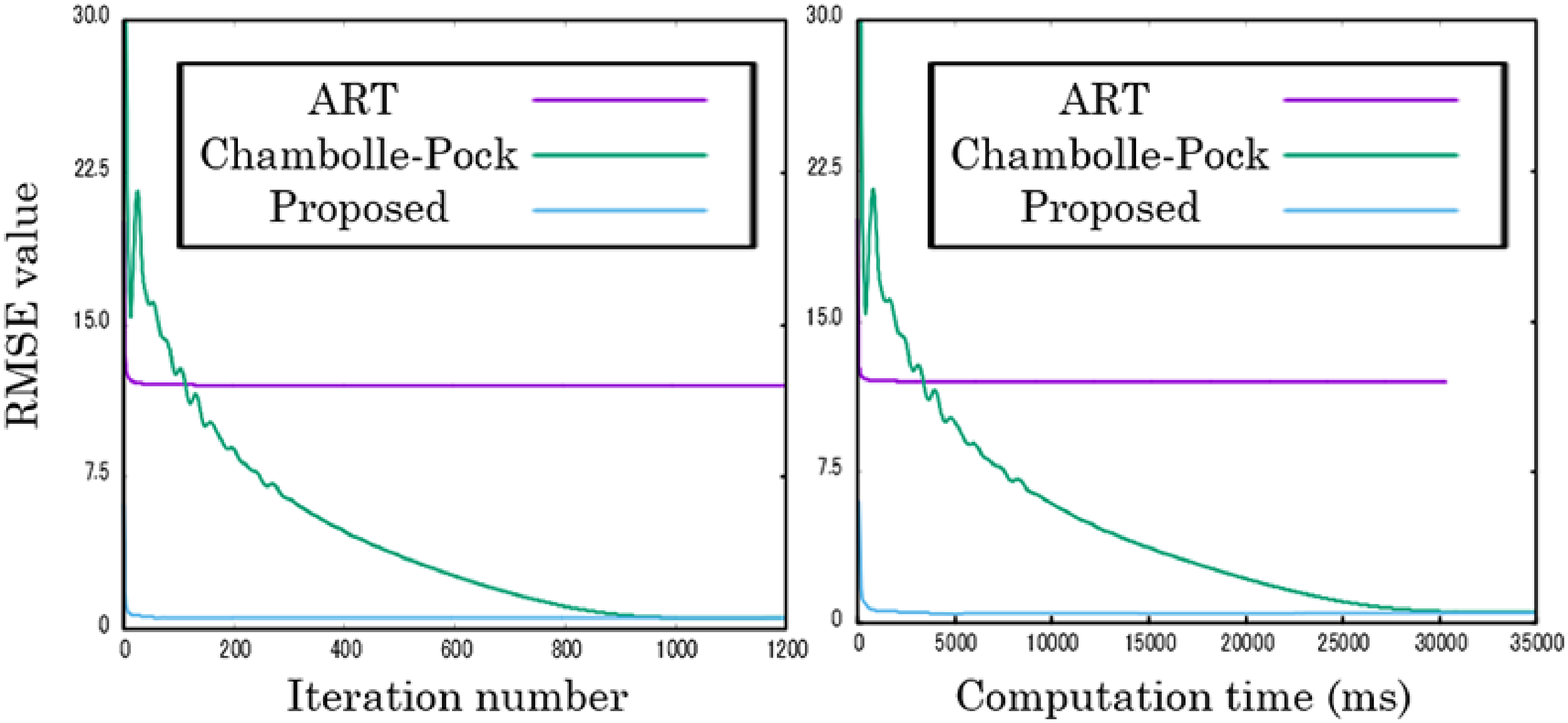}}
\caption{Comparison of convergence properties among Algorithm 2, the Chambolle-Pock algorithm, and the ART algorithm in the few-view CT experiment. Iteration number (computational time) versus RMSE reconstruction error value was plotted.}
\end{figure}
\section{CONCLUSIONS}
\par In this paper, we proposed new image reconstruction algorithms for two typical TV-regularized reconstruction problems in tomography. The first algorithm (Algorithm 1) minimizes the TV-regularized WLS cost function for the low-dose CT, and the second algorithm (Algorithm 2) minimizes the TV penalty under the data fidelity linear constraint for the few-view CT. The both algorithms were derived from the saddle point reformulation of each optimization problem followed by the preconditioning using the ramp filter of FBP reconstruction algorithm and applying the Alternating Projection Proximal (APP) iterative algorithm. We demonstrated that both Algorithm 1 and Algorithm 2 converge very fast while its convergence to the exact solution of each optimization problem is guaranteed.
\par On the low-dose CT reconstruction, very recently, we began to use a shift-variant preconditioner instead of the smoothed ramp filter, which approximates $D=(\tau AA^T+W^{-1})^{-1}$ better, and confirmed an essential improvement. On the few-view CT reconstruction, we began to extend the formulation to $\min_{\vec{x}}\ \psi(\vec{x})\ {\rm subject}\ {\rm to}\ \parallel A\vec{x}-\vec{b}\parallel^2\leq \epsilon^2,\ \vec{x}\geq 0$ by introducing the error tolerance $\epsilon^2$. These newest results will be published elsewhere.
\acknowledgments
This work was partially supported by JSPS KAKENHI Grant Number 15K06103.

\end{document}